\documentstyle[12pt]{article}
\def\c {\tilde{\chi}}
\def\s {\sigma}
\def\q {\bar{q}}
\def\e {\epsilon}
\def\be{\begin{equation}}
\def\ee{\end{equation}}
\def\ba{\begin{array}}
\def\ea{\end{array}}
\def\EQ{\begin{equation}}
\def\EN{\end{equation}}
\def\bea{\begin{eqnarray}}
\def\eea{\end{eqnarray}}
\def\to{\rightarrow}
\def\goto{\longrightarrow}
\def\sa{\hspace{0.1in}}

\def\sb{\hspace{0.2in}}

\def\O{{\cal O}}
\def\F{{\cal F}}
\def\K{{\cal K}}
\def\M{{\cal M}}

\def\T{{\cal T}}

\begin{document}
\oddsidemargin 5mm
\setcounter{page}{0}
\renewcommand{\thefootnote}{\fnsymbol{footnote}}
\newpage
\setcounter{page}{0}
\begin{titlepage}
\begin{flushright}
DAMTP-HEP-94/89
\end{flushright}
\vspace{0.5cm}
\begin{center}
{\large {\bf A method to determine the operator content of perturbed
    conformal field theories}}\\
\vspace{1.5cm}
{\bf
Anni Koubek\footnote{Work supported by PPARC grant no. GR/J20661} }\\
\vspace{0.8cm}
{\em Department of Applied Mathematics and Theoretical Physics,\\
Silver Street,
CB3 9EW Cambridge, UK \\ E-mail: a.koubek@amtp.cam.ac.uk}
\end{center}
\vspace{6mm}
\begin{abstract}
  A method to determine the full structure of the space of local
  operators of massive integrable field theories, based on the form
  factor bootstrap approach is presented.  This method is applied to
  the integrable perturbations of the Ising conformal point. It is
  found that the content of local operators can be expressed in terms
  of fermionic sum representations of the characters $\c(q)$ of the
  Virasoro irreducible representations of the minimal model
  $\M_{3,4}$. The space of operators factorises into chiral components
  as $Z=\sum \c(q) \c(\q)$, but with the relation $\q=q^{-1}$.
\end{abstract}
\vspace{5mm}
\begin{center}
October 1994\\
revised December 1994
\end{center}
\end{titlepage}
\newpage

\section{Introduction}

In this letter we shall present a method in order to determine the
operator content of an integrable massive field theory. It is
based on the form factor bootstrap approach \cite{Karowski,nankai},
which provides
a tool to classify {\em all} local operators of the theory under
consideration. This is possible because the form factor equations are
valid for every local operator of a theory. Therefore determining the
whole solution space of these equations is equivalent to determine the
content of local operators of the theory.

As an example, we shall apply this method mainly to the integrable
perturbations of the critical Ising model. The thermal perturbation
can be described in terms of free massive Majorana fermions, or
equally in terms of a bosonic particle which interacts through an
$S$-matrix $S=-1$. This theory consists of two sectors. One
contains the monomials in the fermion fields, while the other sector
contains fields which aquire a phase $e^{i \pi}$ when moved around the
fermions. The basic fields in this second sector are the order and
disorder fields $\s (x) $ and $\mu(x)$.

The magnetic perturbation is described by a massive field theory
containing 8 particles \cite{Zam}, and is intrinsically related to the
$E_8$ algebra. There are no internal symmetries in this theory and
all operators are expected to be mutually local.

Let us summarise the description of an integrable theory in the form
factor approach.
 We parametrise the momenta of the asymptotic states in
terms of the mass $m$ of the particles and the rapidity variables $\beta_i$
\[
p_i^0\,=\,m \cosh\beta_i\,\,\, , \,\,\, p_i^1\,=\,m
\sinh\beta_i\,\,\, . \]
Form factors are matrix elements of a local operator ${\cal O}$
between the vacuum and the set of asymptotic states, \be \F_n^{\cal O}
(\beta_1,\beta_2,\ldots,\beta_n) \,=\, \langle 0\mid{\cal O}(0,0)\mid
Z( \beta_1), Z(\beta_2),\ldots,Z(\beta_n)\rangle_{in}\sb .
\label{FoF} \ee
Their knowledge determines the
correlation functions which can be expressed as \be\langle
\O(z,\bar{z})\,\O(0)\rangle\,=\label{correlation}\ee
$$ \left (\frac {\bar{z}}{z}\right )^s \sum_{n=0}^{\infty}
\int \frac{d\beta_1\ldots d\beta_n}{n! (2\pi)^n}
\vert F_n^{\cal O}
(\beta_1,\beta_2,\ldots,\beta_n) \vert^2 e^{-mr \sum_i \cosh \beta_i}
\sb ,
$$ where $r$ denotes the radial distance in the Euclidean space, {\em
  i.e.} $r=\sqrt{x_0^2 + x_1^2}$, and $s$ is the spin of the operator
$\O$.

For systems with scalar particles, the functional
equations known as the {\em form-factor axioms} are
\cite{Karowski,nankai, Yurov-Zam}:

\vspace{3mm}

\begin{tabular}{||l|c||}\hline
 & \\
$
\F^\O_{\e_1\dots\e_i\e_{i+1}\dots\e_n}
(\beta_1,\dots,\beta_i,\beta_{i+1},\dots\beta_n) =$ & \\
$ S_{\e_i\e_{i+1}}(\beta_i-\beta_{i+1})\,
\F^\O_{\e_1\dots\e_{i+1}\e_{i}\dots\e_n}
(\beta_1,\dots,\beta_{i+1},\beta_{i},\dots\beta_n) \sb
\label{wat1}$ & (F1)
\\  \hline  & \\
$ \F^\O_{\e_1\e_2\dots\e_n} (\beta_1+2\pi i,\beta_2,\dots,\beta_n )=
e^{2 i \pi (s_{\e_n}+w_{\e_n} +\sum w_{\e_i\e_n})} \F^\O_{\e_2\dots\e_n\e_1}
(\beta_2,\dots,\beta_n,\beta_1 ) \sb
\label{wat2} $ & (F2) \\
\footnotesize
 $s_{\e_n}$ denotes the spin of the particle $\e_n$;
$w_\e$
is the mutual locality index between &  \\ \footnotesize
 $\O$ and $Z_{\e}$; $w_{\e_i\e_n}$
denotes the exchange properties of $Z_{\e_i}$ and $Z_{\e_n}$.
& \\ \hline &  \\
$ \F^\O_{\e_1\dots\e_n}
(\beta_1+\Lambda,\dots,\beta_n+\Lambda ) = e^{s\Lambda}
\F^\O_{\e_1\dots\e_n} (\beta_1,\dots,\beta_n ) \sb \label{prop2} $
 & (F3) \\  \hline
& \\ $
\F_{\e_1,\dots,\e_n}(\beta_1+\Lambda,\dots,\beta_i +\Lambda,
\beta_{i+1}, \dots , \beta_{n}) = O(e^{S_n^i |\Lambda |}) \sb {\rm for}
\sb |\Lambda| \sim \infty \sb
\label{asym}$ & (F4) \\
\footnotesize with  $\max(S_n^i) < \infty$  & \\   \hline & \\
$ -i \lim_{\beta'\to \beta}(\beta'-\beta) \F^\O_{\e\e\e_1\dots\e_n}
(\beta'+i\pi,\beta,\beta_1,\dots,\beta_n) =$ & \\
$ \left
(1-e^{2 i \pi (w_\e+\sum w_{\e_i\e})} \prod_{i=1}^{n}
S_{\e\e_i}(\beta-\beta_i) \right ) \F^\O_{\e_1\dots\e_n}
(\beta_1,\dots,\beta_n ) \sb  \label{kin} $ & (F5) \\  \hline & \\
\footnotesize If the  two-particle scattering amplitude
exhibits a pole with the residue & \\
 $ -i \lim_{\beta'\to
  iu_{\e_i\e_j}^{\e_k}}(\beta-iu_{\e_i\e_j}^{\e_k})
S_{\e_i\e_j}(\beta) =(\Gamma_{\e_i\e_j}^{\e_k})^2 \sb ,
\label{residueS}$ then & \\
$ -i \lim_{\beta'\to \beta}(\beta'-\beta)
\F^\O_{\e_1\dots\e_i\e_j\dots\e_n} (\beta_1,\dots,\beta'+i\bar
u_{\e_i\e_k}^{\e_j},\beta-i\bar u_{\e_j\e_k}^
{\e_i},\dots,\beta_{n-1}) = $ & \\
$ =\Gamma_{\e_i\e_j}^{\e_k}
\F^\O_{\e_1\dots\e_k\dots\e_n} (\beta_1,\dots,\beta,\dots,\beta_{n-1}
) \sb \label{bounds} $ & (F6)\\ & \\ \hline
\end{tabular}

\vspace{3mm}

Further we will use the following notation: We introduce the character
$\chi$ of the conformal minimal model as \be \chi_{r,s}^{(p,q)} =
q^{(h-c/24)} \c_{r,s}^{(p,q)} \sb , \ee where $c$ and $h$ denote the
central charge and the conformal dimension of the primary operator
respectively. Therefore $\c$ encodes the degeneracies in the Virasoro
representation \cite{rocha}, \be \c_{r,s}^{p,q} =
\frac1{(q)_\infty}\sum_{\mu=-\infty}^\infty (q^{\mu (\mu p q + r q - s
  p)}- q^{(\mu p+r)(\mu q+s)})\sb .
\label{chardef} \ee
Further, define $P(m,n)$ to be the number of partitions of $n$ into
numbers whose value do not exceed $m$. These partitions are generated
by \be \frac 1{(q)_m} = \sum_{n=0}^{\infty} P(m,n) q^n
\label{parts}\sb ,
\ee where
\be(q)_m \equiv \prod_{i=1}^m (1-q^i) \sb .\label{GF}\ee

\section{The thermal perturbation of the Ising model}

The theory contains only one particle, and the form factors can be
parametrised as \cite{Yurov-Zam,CM,MTS}
 \be \F_n(\beta_1\dots\beta_n)
= Q_n(x_1,\dots,x_n) H_n (\s_n)^{\frac 12 \delta_{w,0}}\frac
1{(\s_n)^N} \prod_{i<j}^n \tanh \frac {(\beta_i-\beta_j)}{ 2}
\label{pi}\sb , \ee where $\s_i$ denote the elementary symmetric
polynomials \cite{Macdon}, $H_n$ are normalisation constants
and $x_i=e^{\beta_i}$. This parametrisation reduces the recursion
relation (\ref{kin}) to
 \be\ba{ll} Q_{n+2} (-x,x,x_1,\dots,x_n) = x^{2
  N} Q_n (x_1,\dots,x_n)\sa , & w=\frac 12\sb ,
\label{recisingo} \\
Q_{n+2} (-x,x,x_1,\dots,x_n) = 0 \sa , & w= 0\sb . \ea \ee

The integer $N$ in the parametrisation (\ref{pi}) introduces a grading
into the space of operators. Increasing $N$
augments the divergence of the corresponding form factor, if the spin
$s$ of the operator $\O$ is kept fixed. Therefore, in analogy to CFT,
one can define as the chiral operators those with the mildest
ultraviolet behaviour, $N=0$ \cite{CM}.
We have therefore two discrete parameters in the massive theory,
namely the integer $N$ and the spin $s$. The
spin $s$ can be related to the dimensions $(h,\bar{h})$ of the
conformal fields which we obtain taking the UV limit as usual through
$s=h-\bar{h}$. The parameter $N$ has no obvious interpretation in the
conformal limit. Its role will be clarified later on.

\subsection{Operators in the sector $w=\frac 12$}

The sector $w=\frac 12$ contains the order and the disorder fields
$\s$ and $\mu$, and has been investigated in \cite{Yurov-Zam,CM}.  The
form-factors of both operators have been calculated and are given by
$N=0$ and $Q_n=const.$ in (\ref{pi}); further, in \cite{CM} the descendent
operators of $\s(x)$ have been determined.

The method we want to describe does in general not resolve the
recursion relations but rather counts the number of linearly
independent solutions iteratively.  Namely we use the fact that the
number of solutions of the recursion relations at level $n$, {\em
  i.e.} for the form factor $\F_n$, is given
by the number of solutions at level $n-2$ plus the dimension of the
kernel of the recursion relations (\ref{recisingo}).  This amounts in
a simple counting procedure which only involves the comparison of the
degree of the polynomials $Q_n$ and that of the kernel. This has the
advantage that one finds {\em all} solutions to the form factor
equations, and not only those which can be considered as descendent
operators of some non-trivial primary field. In the case of the Ising
model the kernel of the recursion relation (\ref{recisingo}) is given
by \be \K_n = \prod_{i<j}^n (x_i +x_j) \sb .\label{kernel}\ee The
total degree of this function is $deg(\K_n )= \frac 12 n (n-1)$.  A
Kernel solution is consistent for spin $s$ only if $deg(Q_n) \geq
deg(\K_n)$, that is, $Q_n$ can be written as $Q_n = R(\s_i) \K_n$,
where $R(\s_i)$ is an arbitrary combination of elementary symmetric
polynomials of degree $deg(Q_n)-deg(\K_n)$.

 Using this fact and the
generating function (\ref{GF}), one finds that the number of
solutions at general spin $s$, is generated by \be
F_0\equiv\sum_{m,\, odd} \frac{q^{\frac 12 m (m-1)}}{(q)_m} =
\c_{1,2}^{(3,4)} \sb ,\ee
where the exponents in the nominator corresponds to the spin value,
where a new kernel solution enters. Note that this formalism leads  to
fermionic sum expressions \cite{kedem} of the character.

Similarly one can analyse the even form-factors, being related to the
disorder field $\mu$. The counting procedure can be carried out
analogously, and one finds that the number of chiral operators for
spin $s$ is generated by \be G_0\equiv\sum_{m,\, even} \frac{q^{\frac
    12 m (m-1)}}{(q)_m} = \c_{1,2}^{(3,4)} \sb .\ee As it is well
known, both order and disorder fields are related to the conformal
operator $\phi_{1,2}$.

As a next step we investigate solutions corresponding to higher $N$.
Increasing $N$ we can gradually build up the whole space
of operators in the massive theory.  The number of operators at level
$N$ is generated by $$ F_N\equiv \sum_{m=1,odd}^{\infty}
\frac{q^{\frac 12 m (m-1)-N m}}{(q)_m} \sb $$ in the odd sector, and
by
$$ G_N\equiv \sum_{m=0,even}^{\infty} \frac{q^{\frac 12 m (m-1)-N
    m}}{(q)_m} \sb $$ in the even sector.  These functions satisfy the
recursion relations \be F_N = F_{N-1} - \q^N G_{N-1}\sa , \sb G_N =
G_{N-1} + \q^N F_{N-1}\sb , \ee with the initial conditions $G_0 = F_0
= \c_{1,2}(q)$,
 which can easily be solved to give $F_N = G_N =
\prod_{k=1}^N (1+\q^k) F_0$.
The whole space of states is
generated by
\be\lim_{N\to\infty} F_N =
 \c_{1,2} (\q) \c_{1,2}(q) \sb ,\label{chir}\ee and
similarly for $G_\infty$.

Let us summarise the key features of the method described, which
generalise also to other integrable massive field theories \cite{a14}:
\begin{itemize}
\item The content of chiral operators is given by fermionic sum
  expressions of the characters of the corresponding Virasoro
  irreducible representation (or of sums of the characters).
\item The whole space of operators is constructed by taking the limit
  $N\to \infty$. It can be formally decomposed into conformal
  characters as $\sum \c(q) \c(\q)$ where the variables $q$ and $\q$
  are {\em not} independent as in CFT but $\q = q^{-1}$.
\item Finally, one obtains for finite
  $N$ some `finitized' expressions for the Virasoro characters. In
  this approach they appear naturally in the counting procedure of
  states.  Note that the same expressions turn up in the study of
  corner transfer matrices where the finitization is related to the
  fact that one has a discrete system of finite size (see {\em e.g.}
  \cite{bc} for the thermal perturbation of the Ising model). It would
  be interesting to understand whether there is a deeper reason for
  this relation.
\end{itemize}

\subsection{Operators in the sector $w=0$}

For the sector $w=0$ the kinematical recursion relation maps the form
factors onto zero, $\F_n(\beta+i\pi,\beta,\dots) =0$.  It follows that
form factors with different particle number $n$ are not linked, and
therefore {\em any} kernel solution will represent an acceptable form
factor from the point of view of the form factor equations.

We start our analysis with the space related to the even form factors.
The fundamental operator is the energy density, whose form factor is
given by $\F_2 = \sinh \frac \beta 2$. Comparing this with the
parametrisation (\ref{pi}), we find that in this sector the primary
operator corresponds to a solution with $N=1$.

Using the counting method as before, we find that the solutions at
level $N$ are generated by
\be f_N \equiv \sum_{n,even}
\frac{q^{\frac{n^2}2-Nn}}{(q)_n} \sb .\ee
Expressing the lowest terms in Virasoro characters, one finds that
$f_0=\c_{1,2}(q)$ and $f_1$, which we conjectured to generate the
chiral operators, is given by
$f_1 = \c_{1,1}(q) + \c_{1,3}(q)$, as expected.

Since the form factors are not linked by the recursive equations for
every operator in this sector there is only one contribution in the
sum in (\ref{correlation}), and the critical exponents can be
calculated explicitly.  One can show that all operators corresponding
to solutions $Q_n$ in the space $N=0$, are operators which will scale
in the ultraviolet limit as elements of the identity module, while the
operators in the $N=1$ space form the space of descendents of the primary field
$\phi_{1,3}$ \footnote{The operators of the space $N=0$ appear in the
  space $N=1$ as solutions $Q_n \sim \s_n$.}.

In order to find the decomposition of this expression into characters
let us introduce \be g_N \equiv \sum_{n,odd} \frac{q^{\frac {n^2-1}{2}
    - N n}}{(q)_n} \sb .
\label{gen3} \ee
These generating functions satisfy the recursion identities \be\ba{l}
f_N = f_{N-1} + \q^{N-1} g_{N-1} \sb , \\ g_N = g_{N-1} + \q^N f_{N-1}
\sb , \ea \ee with initial conditions $f_0 = \c_{1,1}$ and
$g_0=\c_{1,3}$.  Their solution is given by \cite{abf,melzer}
\be\ba{l} f_N = D_N^{(1)}(\q) \c_{1,1} + D_N^{(2)} (\q) \c_{1,3} \\
g_N = \q D_N^{(2)}(\q) \c_{1,1} + D_N^{(1)} (\q) \c_{1,3} \ea \ee
where \bea D_N^{(1)}(\q) &=& \sum_{\mu=-\infty}^{\infty} \q^{ \mu (12
  \mu +1)} \left [ \ba{c} 2 N \\ N-4 \mu \ea \right ]- \q^{(3 \mu+1)
  (4 \mu +1)} \left [ \ba{c} 2 N \\ N-1-4 \mu \ea \right ] \nonumber
\sb ,\\ D_N^{(2)}(\q) &=& \sum_{\mu=-\infty}^{\infty} \q^{ \mu (12 \mu
  +1)} \left [ \ba{c} 2 N \\ N-1-4 \mu \ea \right ]- \q^{(3 \mu+1) (4
  \mu +2)} \left [ \ba{c} 2 N \\ N-2-4 \mu \ea \right ] \sb ,\eea
and $ [\ba{c} a\\b\ea ] =\frac{(q)_a}{(q)_{a-b}(q)_b}$ denote the
Gaussian polynomials.  These expressions satisfy \be D_N^{(1)}(\q)
\stackrel{N\to\infty}{\goto} \c_{1,1}(\q) \sa ,\sb D_N^{(2)}(\q)
\stackrel{N\to\infty}{\goto} \c_{1,3}(\q) \sb ,\ee which means that
for $N \to \infty$ the space of local operators in this sector can be
written as \be f_{\infty} =\c_{1,1}(\q) \c_{1,1}(q) + \c_{1,3} (\q)
\c_{1,3}(q) \sb . \ee

Finally we analyse the odd form factors in the sector $w=0$.  From the
parametrisation (\ref{pi}) we find that form-factors of the operators
with the mildest ultraviolet behaviour ($N=0$) will have half integer
spin.  We choose as our basic operator the fermion, whose form factor
is given by $\F_1 = \s_1^{\frac 12} =e^\frac\beta 2$.  Carrying out a
similar analysis as for the even form factors one finds that the
generating function for the chiral operators is given by $ g_0 =
\c_{1,3} $, where $G_N$ is given in (\ref{gen3}).  One can easily
check that the ultraviolet dimensions are those of the corresponding
conformal operators.

The whole space of operators in this sector is generated by \be
g_\infty = \c_{1,1}(\q) \c_{1,3}(q) + \q \c_{1,3}(\q) \c_{1,1}(q)\sb .
\ee The space of operators has the same structure as the space of
descendent operators of the conformal operators $(\frac 12 ,0)$ and
$(0,\frac 12)$.

Summarising, we found the descendent spaces of the fermions, the energy
density and the identity operator in this sector. All are expressed in
terms of the characters of the critical Ising model. It is interesting
to note that the counting method selects automatically a base in the
space of operators which is isomorphic to the corresponding Virasoro
irreducible representation.

\section{The magnetic perturbation}

The perturbation of the critical Ising model by the operator
$\phi_{1,2}$ couples the model to an external magnetic field at
critical temperature. This system develops a finite correlation length
and hence is massive.
The on-shell structure of the theory has been determined in
\cite{Zam}, and contains 8 scalar massive
particles. The $S$-matrix of the fundamental particle is given by
$$S_{11}(\beta_1-\beta_2) = f_{2/3}(\beta_1-\beta_2)f_{2/5}
(\beta_1-\beta_2)f_{1/15}(\beta_1-\beta_2)$$ with \be
f_\alpha(\beta) \equiv \frac{\tanh \frac 12(\beta+i\alpha\pi)}{\tanh
  \frac 12 (\beta-i\alpha\pi)} \sb .\ee All other $S$-matrix elements can
be calculated by using the bootstrap equations. Because of the lack of
space we are not able to give a detailed account of our
investigations. We will present the major results, in order to show
how the counting method can be applied also to relatively complicated
systems, and give a complete presentation elsewhere \cite{a14}.

We start our analysis by applying the counting method to the
fundamental particle and include the others by using the bound state
equation (F6). In order to parametrise the form-factor conveniently, we
introduce
  \be \zeta(\theta,\alpha) \equiv
 \exp \left \{
\int_0^\infty \frac{dx}x \frac{\cosh x (\frac 12 - \alpha)}
{\cosh \frac x2}
\frac{\sin ^2 \frac{ x(i \pi-\beta)}{2\pi}}{\sinh x} \right \}
\sb ,\ee
and
$$ <\alpha> \equiv \sinh \frac 12 (\beta + i \pi\alpha)
\sinh \frac 12 (\beta - i \pi\alpha) \sb .$$

The form-factors corresponding to $n_1$ particles of type 1 can then
be parameterised as \be \F_{n_1} (\beta_1,\dots,\beta_{n_1}) =
\label{pe8} \ee
$$
H_{n_1} Q_{n_1}(x_1,\dots,x_n) \frac{1}{\s_{n_1}^{N_1}}
 \prod_{i<j}^{n_1} \frac{\zeta(\beta_{ij},\frac 23)\zeta(\beta_{ij},\frac
25)\zeta(\beta_{ij},\frac 1{15})}{(x_ix_j)^{\frac 32}
<1>^{\frac 12} <\frac 23> <\frac
25> <\frac 1{15}>}$$ where $\beta_{ij} = \beta_i -\beta_j$.
 Note that this parametrisation is an analytic continuation of the
parametrisation of the form-factors of the particle of the
Bullogh-Dodd model \cite{FMS-BD}.  The degree of the
kernel of the recursion relations is given by $deg(\K_{n_1}) =\frac 72
n_1 (n_1-1)$, while the degree of $Q_{n_1}$ is $deg(Q_{n_1}) = \frac
32 n_1 (n_1-1)$, which leads to the $q$-sum expression
\be
\T = \sum_{m_1}
\frac{q^{2 m_1 (m_1-1)}}{(q)_{m_1}}\sb .\label{exprr}\ee
In order to determine the contributions to this $q$-sum expression
arising from particle $2$, we apply the bound
state equation to (\ref{pe8}). For the form factors with $n_1$
particles $1$, and $n_2$ particles $2$ we find
$$
\F_{1,\dots,1,2\dots,2}=H_{n_1,n_2}
\frac {1}{\left ( \s_{n_1}(x_1,\dots,x_{n_1})\right )^{N_1}\left
  (\s_{n_2}
(x_{n_1+1},\dots,x_{n_1+n_2})\right )^{N_2}}\sa \times $$
\be\times Q_{n_1,n_2}(x_1,\dots,x_{n_1};x_{n_1+1},\dots,x_{n_1+n_2}) \times
\label{f12}
\ee
$$\times  \prod_{i<j}^{n_1} \frac{\zeta(\beta_{ij},\frac
23)\zeta(\beta_{ij},\frac
25)\zeta(\beta_{ij},\frac 1{15})}{(x_ix_j)^{\frac 32}
<1>^{\frac 12} <\frac 23> <\frac
25> <\frac 1{15}>}\sa \times$$
$$\times\prod_{i=1}^{n_1} \prod_{i=1}^{n_2} \frac{1}{(x_i
x_j)^3} \frac{\zeta(\frac 7{15})\zeta(\frac 3{5})\zeta(\frac
1{5})\zeta(\frac 4{15})
}{<\frac{4}{5}><\frac{3}{5}><\frac{4}{15}><\frac{13}{15}><\frac{7}{15}>}
\sa \times
$$
$$ \times \prod_{i<j}^{n_2} \frac{1}{(x_i
x_j)^6} \frac{<0> \zeta(\frac 2{3})\zeta(\frac 4{15})\zeta(\frac
1{5})\zeta(\frac 2{5})^2\zeta(\frac 7{15})\zeta(\frac 1{15})
}{<1><\frac{2}{3}><\frac{4}{5}><\frac{7}{15}><\frac{1}{15}>
<\frac{4}{15}>
<\frac{3}{5}><\frac{2}{5}><\frac{16}{15}><\frac{2}{3}>}
\sb .
$$
Comparing the dimension of the functions $Q_{12}$ and $Q_{22}$ with the
dimension of the respective kernels, one finds that the exponent due to
interaction of particles 1 and 2 is $2 m_1^2 +4 m_2^2 +  4 m_1 m_2-2
m_1 - 4 m_2$, modifying the
expression (\ref{exprr}) to \be
\T = \sum_{m_1,m_2} \frac{q^{\left \{ (m_1,m_2)\left (\ba{cc} 2 & 2\\ 2& 4\ea
\right ) \left ( \ba{c} m_1 \\m_2\ea \right )- 2 m_1 -4 m_2\right \} }}
{(q)_{m_1}(q)_{m_2}}\sb .\label{exprr2}\ee

Notice that the pole structure in (\ref{f12}) is not minimal (in the
sense that they exhibit poles also in non-physical locations). This is
 no obstacle
for the counting method since only the {\em difference} between the
degrees of the functions $Q$ and the kernels $\K$ enters.

\begin{figure}
\begin{picture}(300,100)(-100,-30)
\multiput(0,0)(30,0){7}{\circle*{5}}
\put(0,0){\line(1,0){180}}
\put(120,0){\line(0,1){30}}
\put(120,30){\circle*{5}}
\put(0,-8){\makebox(0,0){$1$}}
\put(30,-8){\makebox(0,0){$3$}}
\put(60,-8){\makebox(0,0){$5$}}
\put(90,-8){\makebox(0,0){$7$}}
\put(120,-8){\makebox(0,0){$8$}}
\put(150,-8){\makebox(0,0){$6$}}
\put(180,-8){\makebox(0,0){$2$}}
\put(115,30){\makebox(0,0){$4$}}
\end{picture}
\caption{The $E_8$ Dynkin diagram, with the labelling of the nodes used
  in the definition of the generating function (\protect\ref{deft}). }
\label{fig-one}
\end{figure}
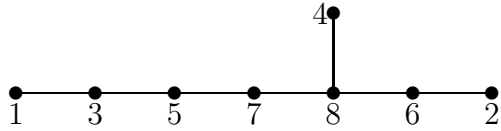

Similarly it is possible to carry out the bootstrap for all particles.
The corresponding expression of the form-factors becomes quite
lengthly, and will be omitted. It will contain factors
$1/\s_{n_i}^{N_i}$, $i=1,\dots,8$, corresponding to each of the 8
particles.  The generating functions for levels $N_1,\dots,N_8$ can be
found to be \be \T_{N_1,N_2,\dots,N_8} \equiv \T_N = \sum_M \frac{q^{M
    C^{-1} M - N C^{-1} M}}{(q)_{m_1}\dots (q)_{m_8}} \label{deft}\ee
where $M$ denotes the multi-index $M=m_1,\dots m_8$ and $C^{-1}$ is
the inverse Cartan matrix of the algebra $E_8$. The relevant labelling
of the nodes of the $E_8$ Dynkin diagram is depicted in fig.
\ref{fig-one}.  The linear terms in the $q$ sum
expressions have been chosen in order to be consistent with the bootstrap.

The lowest of these functions is $\T_0 = \c_{1,1}(q)$ \cite{kedem,wp},
corresponding to the most relevant operators,
while the chiral operators are given by \cite{kedem}
$$\T_{0,1,0,0,0,0,0,0} = \c_{1,1}(q) +\c_{1,2}(q) +\c_{1,3}(q)\sb .$$
Similar higher functions can be expressed in terms of characters as
for example
\bea
  \T_{0,0,1,0,0,0,0,0} &=& \c_{1,1}(q)
 + (1+\q) \c_{1,2} (q) +\c_{1,3}(q) \nonumber \\
 \T_{0,0,0,1,0,0,0,0} &=& \c_{1,1}(q)
+(1+\q) \c_{1,2}(q) +(1+\q)\c_{1,3}(q)\nonumber \\
\T_{0,0,0,0,1,0,0,0} &=& (1+\q^2)\c_{1,1}(q) +(1+\q+\q^2)
\c_{1,2}(q) +(1+\q+\q^2)\c_{1,3}(q) \sb .\eea
These expressions correspond to the scalar partition function of the
Ising model in its lowest orders in $\q$.

The generating functions (\ref{deft}) satisfy the recursion identities
$$\T_{N+\delta_{N_i}} = \T_{N- C \delta_{N_i}} + q^{C^{-1} N}
\T_{N-\delta_{N_i}} \sb ,$$ which allow to reconstruct the space of
operators for higher levels.  We have carried out extensive numerical
calculations which show that for $N \to \infty$ the functions $\T_N$
approach the scalar partition function \be Z = \c_{1,1}(q)
\c_{1,1}(\q)+ \c_{1,2}(q) \c_{1,2}(\q) +
 \c_{1,3}(q) \c_{1,3}(\q) \ee as expected.
It would be interesting to find exact solutions for these recursion
relations, and determine their relation to the finitized characters
for this model, which arise from carrying out the Bethe ansatz to the
dilute $A_3$-model \cite{wp}.

 \section{Conclusions}

We presented a method in order to determine the full content of local
operators of an integrable massive field theory. We have discussed its
application to the integrable perturbations of the critical Ising
model. It can be applied to any massive integrable theory, and in
\cite{a14} we will discuss more examples explicitly.

The method consists of a simple counting procedure and leads to
fermionic sum expressions of the characters, or of sums of the
characters. Further, by increasing the possible divergence of the
form-factors, the whole space of operators can be obtained. We found
that this space can be formally factorised into characters of the
corresponding Virasoro irreducible representation of the form
$\sum \c (q) \c(\q)$. This is a quite remarkable result, since the
chiral sectors are not independent in the massive model. This fact is
reflected by the relation $\q = q^{-1}$.

Though the structures of the conformal and massive theories are quite
similar, there is one important difference. The grading in the space of
conformal operators is introduced through the conformal weights of the
operators, $h$ and $\bar{h}$, while in the massive model it is given
through the spin $s$ and the parameters $N$. While the spin can be
linked to the conformal weights as $s=h-\bar{h}$ the parameters $N$
lack such a direct interpretation in terms of the CFT weights. It
is our belief that understanding this connection of $N$ to the Virasoro
structure in the ultraviolet limit, will give an indication
of the algebraic
structure determining the space of local operators in the massive
theory.

\end{document}